\documentclass{aastex6}

\shorttitle{Plasmoid chains growth during viscoresistive evolution of the tilt instability}
\shortauthors{Baty et al.}

\usepackage{bm}%
\usepackage{graphicx}%
\usepackage{amsmath}

\def\ltsima{$\; \buildrel < \over \sim \;$}
\def\gtsima{$\; \buildrel > \over \sim \;$}
\def\simlt{\lower.5ex\hbox{\ltsima}}
\def\simgt{\lower.5ex\hbox{\gtsima}}

\begin{document}

\title{On the growth rate of plasmoid chains during nonlinear viscoresistive evolution of the tilt instability}

\email{hubert.baty@unistra.fr}

\author{ Hubert BATY}
\affiliation{Observatoire Astronomique de Strasbourg, Universit\'e de Strasbourg, CNRS, UMR 7550, 11 rue de l'Universit\'e, F-67000 Strasbourg, France}

\begin{abstract}

We investigate by means of two-dimensional incompressible magnetohydrodynamic (MHD) numerical simulations,
the onset phase of the fast collisional magnetic reconnection regime that is supported by the formation of plasmoid chains
when the Lundquist number $S$ exceeds a critical value. The present study extends previous results obtained
at magnetic Prandtl number $P_m =1$ \citep{bat20} to a range of different $P_m$ values. We use FINMHD code
where a set of reduced visco-resistive MHD equations is employed to form two quasi-singular current layers as
a consequence of the tilt instability. The results reinforce the conclusion that, 
a phase of sudden super-Alfv\'enic growth (when $P_m$ is not too high) of plasmoid chains is obtained,
following a previous quiescent phase during current sheet formation on a slower Alfv\'enic time scale.
We compare our results with predictions from the general theory of the plasmoid instability.
We also discuss the importance of this onset phase to reach the ensuing stochastic time-dependent reconnection regime,
where a fast time-averaged rate independent of $S$ is obtained.
Finally, we briefly discuss the relevance of our results to explain the flaring activity in solar corona 
and internal disruptions in tokamaks.

\end{abstract}


\section{Motivation}

Magnetic reconnection is believed to be the underlying mechanism that explains explosive events
observed in many magnetically dominated plasmas. This is for example the case for flares in the solar corona, or
sawtooth crashes in tokamaks. It is a process of topological rearrangement of magnetic field lines that
can convert a part of the magnetic energy into kinetic energy and heat \citep {pri00}.
However, the timescales involved
in classical two-dimensional (2D) reconnection models within the macroscopic magnetohydrodynamic (MHD) regime are too slow to match the observations or experiments.
Indeed, the reconnection rate predicted by Sweet-Parker (SP) model which scales like $S^{-1/2}$ ($S$ being the
Lundquist number defined as $S = L V_A / \eta$,  where $L$ is the half-length of the current sheet, $V_A$ is the Alfv\'en speed based
on the magnetic field amplitude in the upstream current layer, and $\eta$ is the resistivity), is too low by a few (or even many) orders of magnitude
for the relevant Lundquist numbers \citep {swe58, par57}.
For example, for typical parameters representative of the solar corona, $S$ is of order $10^{12}$, leading to a normalized reconnection rate of order
$10^{-6}$ much lower than the value of $10^{-2} -10^{-1}$ required to match the observations. Furthermore, SP theory assumes a
steady-state process that cannot explain the impulsive (thus even faster) onset phase preceding the main one.

However, it has been realized in the last decade that, even in a magnetofluid approach, a new solution with a rate that is (possibly) fast
enough and almost independent on $S$ can be obtained, provided that $S$ is higher than a critical value of order $10^4$.
This new regime is supported by the formation of plasmoid chains disrupting the current sheet in which they are born, as
obtained in many numerical experiments \citep {sam09, bha09, hua10}.
More precisely, these plasmoids are small magnetic islands due to tearing-type resistive instabilities, constantly forming, moving,
eventually coalescing, and finally being ejected through the outflow boundaries. At a given time, the system appears as an aligned layer structure of plasmoids of
different sizes, and can be regarded as a statistical steady state with a time-averaged reconnection rate that is nearly (or
exactly) independent of the dissipation parameters \citep {uzd10, lou12}.
The modal linear theory of plasmoid instability is based on a preformed static
(i.e. reconnection flows effects are neglected) unstable SP current sheet with a half-width $a \simeq L S^{-1/2}$ \citep {lou07}.
Among the spectrum of many unstable modes (as $k a  \leq 1$ is required if we assume a Harris-type current layer profile having a hyperbolic tangent magnetic field reversal),
the linearly dominant wavenumber $k_p$ follows $k_p L  \simeq 1.4$ $(1 + P_m)^{-3/16} S^{3/8}$ (where $P_m = \nu/\eta$ is the magnetic Prandtl number, i.e. the ratio of
the viscosity coefficient $\nu$ to the resistivity one $\eta$) with a corresponding maximum linear growth rate $\gamma_p$ scaling as $\gamma_p  \tau_A  \simeq 0.62$  $(1 + P_m)^{-5/8} S^{1/4}$,
where $\tau_A = L/V_A$ is the Alfv\'en time based on the current sheet half-length \citep {com16, hua19}.

Beyond these above well admitted results and despite many published papers on the subject, there is no clear consensus
on a theoretical view for the plasmoids-reconnection regime including the onset phase.

The paradoxal result of infinite linear growth rate (see scaling law just above) in an ideal MHD plasma (i.e. infinite $S$) being incompatible with the frozen-in condition that makes
reconnection impossible, an issue has been proposed by considering unstable current layers having a critical aspect ratio $L/a \simeq S^{ \alpha}$, that is
smaller than SP value in the high $S$ limit as $0.25 < \alpha < 0.5$ \citep {puc14}. In this way, the linear growth rate becomes Alfv\'enic and independent of $S$.
The value of the exponent $\alpha$ depends on the current profile \citep {puc18}.
For example,  $\alpha=1/3$ is found for the standard Harris current profile, leading to 
$\gamma_p \tau_A  \simeq 0.62$ (case of zero viscosity)
with the corresponding linearly dominant wavenumber $k_p$ following
the relation $k_p L  \simeq 1.4$ $S^{1/6}$. These results have been confirmed by numerical simulations of preformed static current layers having the correct aspect ratio value, and
seem to remain true when extended to macroscopic current sheets of fixed length that are artificially forced to collapse 
asymptotically towards $a/L \sim S^{-1/3}$ and $a/L \sim S^{-1/2}$
on a time scale of order of $\tau_A$ \citep {ten15, ten16}.  

On the other hand, a second theoretical issue has been proposed by \citet {com16b, com17}
by investigating the plasmoid instability in a dynamically evolving (exponentially shrinking in time and reaching asymptotically a SP aspect ratio)
current sheet. Without any assumption on the critical current sheet aspect ratio for disruption onset, they employ a principle of least time to 
derive it as well as the corresponding dominant mode and associated growth rate. The main difference compared to the approach proposed by the first issue,
is that the dominant mode is not necessarily the linearly fastest one (obtained from a classical static stability study), but the mode that is able to emerge first at the
end of the linear phase. In this way, new scalings that are not simple $S$-power laws
are obtained. For example, the dominant mode growth rate is predicted to follow
a transition between the previous scaling $\gamma_p \propto S^{1/4}$ (for moderate $S$ values, $S \simgt S_c$) and an asymptotic (for very high $S$ values,
$S \gg S_c$) new scaling with a decreasing logarithmic dependence (see Equation 19 in \citet {com16b}, and Equation 32 in \citet {com17}).  
The growth rate can in principle easily attain super-Alfv\'enic values  $\gamma_p  \tau_A  \sim 10 - 100$, 
while remaining finite in the infinite $S$ limit. The precise value of the growth rate and of the corresponding wavenumber also depend on other parameters than $S$,
that are the characteristic time scale of the current sheet formation, the thinning process, the magnetic Prandtl number, and the noise of the system.

This second issue seems to be partly supported by recent 2D numerical MHD simulations, where the coalescence instability between
two parallel currents is chosen as the initial setup providing the thinning process to form the current sheet \citep {hua17}. Indeed, a scaling law transition
is effectively observed, and maximum growth rates with $\gamma_p \tau_A \simeq 10-20$ 
are obtained that are substantially smaller than values predicted by the theoretical model.
The remaining differences between the simulations and the analytical model of
\citet {com16b, com17} are explained by taking into account the effects of the reconnection outflow in a phenomenological model \citep {hua19}.
In our previous study using a different setup  \citep{bat20} (hereafter denoted as Paper I), namely using the  tilt instability between two repelling
antiparallel currents \citep {ri90}, a similar conclusion was drawn with obtained maximum super-Alfv\'enic growth rates $\gamma_p \tau_A \simeq 10$.

Conversely, as the first theoretical model proposed by \citet {puc14} predicts constant and smaller 
growth rates, more precisly with $\gamma_p \tau_A  \sim 1$, it consequently seems to fail to explain these numerical simulations based
on coalescence/tilt setups.
However, when submitting the results of Paper I, a controversial point arises about validity of the diagnostic (i.e. the maximum current density)
used to estimate the growth rate at which plasmoids can grow. 
In the present work, we thus focus on this onset phase leading to the disruption of the current sheets by the formation of many
plasmoids. Using the same MHD code  (FINMHD, \citet {bat19}) and numerical procedure with the tilt instability setup,
we extend the results obtained in Paper I at $P_m = 1$, for a range of different $P_m$ values.
The ensuing statistical steady state with a fast reconnection rate is beyond the scope of the present paper and is left to a future work.
The outline of the paper is as follows. In section 2, we present the MHD code and the initial setup for tilt instability.
Section 3 is devoted to the presentation of the results. Finally, we conclude in section 4.

\section{The MHD code and initial setup}

\subsection{FINMHD equations}
For FINMD, a set of reduced MHD equations has been chosen corresponding to a 2D incompressible model. However,
instead of taking the usual formulation with vorticity and magnetic flux functions for the main variables, another choice
using current-vorticity ($J-\omega$) variables is preferred because of its more symmetric formulation, facilitating the numerical
matrix calculus. The latter choice also cures numerical difficulty due to the numerical
treatment of a third order spatial derivative term \citep{phi07}. To summarize, the following set of equations is (see also \citet{bat19} for more details),
\begin{equation}  
      \frac{\partial \omega}{\partial t} + (\bm{V}\cdot\bm{\nabla})\omega = (\bm{B}\cdot\bm{\nabla})J + \nu \bm{\nabla}^2 \omega ,
\end{equation}
\begin{equation}      
        \frac{\partial J }{\partial t} + (\bm{V}\cdot\bm{\nabla})J =  (\bm{B}\cdot\bm{\nabla})\omega + \eta \bm{\nabla}^2 J +  g(\phi,\psi) ,
\end{equation}
\begin{equation}                 
     \bm{\nabla}^2\phi = - \omega ,
 \end{equation}
\begin{equation}                        
     \bm{\nabla}^2\psi = - J ,  
\end{equation}
with $g(\phi,\psi)=2 \left[\frac{\partial^2 \phi}{\partial x\partial y}\left(\frac{\partial^2 \psi}{\partial x^2} - \frac{\partial^2 \psi}{\partial y^2}\right) - \frac{\partial^2 \psi}{\partial x\partial y}\left(\frac{\partial^2 \phi}{\partial x^2} - \frac{\partial^2 \phi}{\partial y^2}\right)\right]$.
As usual, we have introduced the two stream functions, $\phi (x, y)$ and $\psi (x, y)$, from the fluid velocity $\bm{V} = {\nabla} \phi \wedge \bm{e_z}$ and magnetic field $\bm{B} = {\nabla} \psi \wedge \bm{e_z}$ ($\bm{e_z}$
being the unit vector perpendicular to the $xOy$ simulation plane).
$J$ and vorticity $\omega$ are the $z$ components of the current density and vorticity vectors, as $\bm{J} = \nabla \wedge \bm{B}$ and $\bm{ \omega} = \nabla \wedge \bm{V}$ respectively (with units using $\mu_0 = 1$).
Note that we consider the resistive diffusion via the $\eta \bm{\nabla}^2 J $ term ($\eta$ being assumed uniform for simplicity), and also a viscous term
$\nu \bm{\nabla}^2 \omega$ in a similar way (with $\nu$ being the viscosity parameter also assumed uniform).
The above definitions results from the choice $\psi \equiv A_z$, where $A_z$ is the $z$ component of the potentiel vector $\bm{A}$ (as $\bm{B} = \nabla \wedge \bm{A}$). This choice
is the one used in \citet{ng07} or in \citet{bat16},
and different from the one used by \citet{lan07} where the choice $\psi \equiv - A_z$ is done.
In the latter case, the two Poisson equations (i.e. Equations 3-4) involve an opposite sign in the right hand sides. Note that thermal pressure gradient is naturally absent from our set of equations. 
Note also that, an advantage of the above formulation over a standard one using the velocity and magnetic field vectors ($\bm{V}, \bm{B}$)  as
the main variables, is the divergence-free property naturally ensured for these two vectors.

\subsection{FINMHD numerical method}

Simulating the mechanism of magnetic reconnection in the high Lundquist number regime requires the use of
particularly well adapted methods. Conventional codes generally lack some convergence properties
to follow the associated complicated time dependent bursty dynamics \citep{kep13}. 
Despite the fact that they are not commonly used, finite element techniques allows to treat the early formation of
quasi-singularities \citep{str98, lan07}, and the ensuing magnetic reconnection in an efficient way \citep{bat19}.

FINMHD code is based on a finite element method using triangles with quadratic
basis functions on an unstructured grid. A characteristic-Galerkin scheme is chosen 
in order to discretize in a stable way the Lagrangian derivative $\frac{\partial  }{\partial t} + (\bm{V}\cdot\bm{\nabla}) $ appearing
in the two first equations \citep{bat19}.
Moreover, a highly adaptive (in space and time) scheme is developed in order to follow the rapid
evolution of the solution, using either a first-order time integrator (linearly unconditionally stable) or a second-order one
(subject to a CFL time-step restriction). Typically, a new adapted grid can be computed at each time step, by searching the grid
that renders an estimated error nearly uniform. The finite elements Freefem++ software allows to do this  \citep{hec12}, by using
the Hessian matrix of a given function (taken to be the current density in this study).
The technique used in FINMHD has been tested on challenging tests, involving 
unsteady strongly anisotropic solution for the advection equation, formation of shock structures
for viscous Burgers equation, and magnetic reconnection for the reduced set of MHD equations.
The reader should refer to \citet{bat19} for more details.

\subsection{The initial setup}

The initial magnetic field configuration for tilt instability is a dipole current structure similar to the dipole vortex flow pattern in
fluid dynamics \citep{ri90}.
It consists of two oppositely directed currents embedded in a constant magnetic field.
Contrary to the coalescence instability based on attracting parallel current structures, the two antiparallel currents in the configuration tend to repel.
The initial equilibrium is thus defined by taking the following magnetic flux distribution,
\begin{equation}
    \psi_e (x, y)=
    \left\{
      \begin{aligned}
        &\left(\frac{1}{r} - r\right)\frac{y}{r} ~~~& & if ~~ r > 1 , \\
        &-\frac{2}{\alpha J_0(\alpha)}J_1(\alpha r)\frac{y}{r} ~~~& & if ~~ r\leq1 .\\
      \end{aligned}
      \right.
  \end{equation}
  
 And the corresponding current density is,
       \begin{equation} 
    J_e (x, y) =
    \left\{
      \begin{aligned}
        &~~~~~~~~~~~~0 ~~~& & if ~~ r > 1 , \\
        &-\frac{2\alpha}{J_0(\alpha)}J_1(\alpha r)\frac{y}{r} ~~~& & if ~~ r\leq1 ,\\
      \end{aligned}
    \right.
\end{equation}

\noindent where 
$r=\sqrt{x^2+y^2}$, and $J_0$ et $J_1$  are Bessel functions of order $0$ and $1$ respectively.
Note also that $\alpha$ is the first (non zero) root of $J_1$, i.e. $\alpha = 3.83170597$.
\medskip
\begin{figure}
\centering
 \includegraphics[scale=0.8]{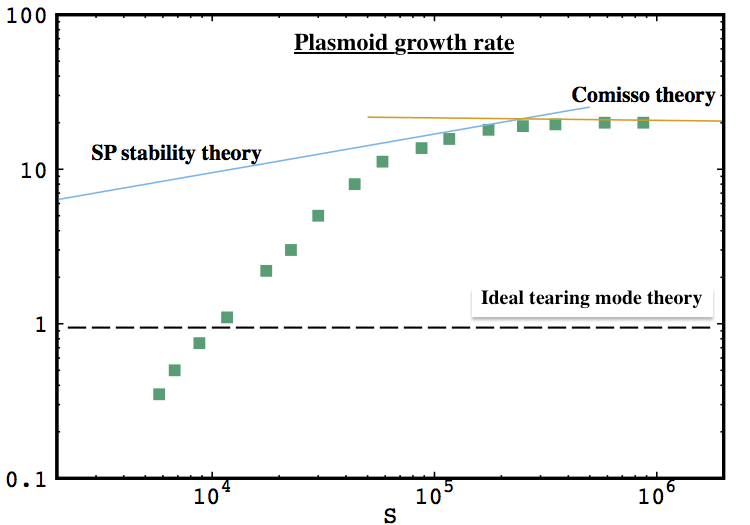}
  \caption {Growth rate $ \gamma_p  t_A$ obtained for plasmoid formation in simulations at different $S$ for $P_m = 1$
  (squares), expected from SP linear theory scaling as $0.9 \times S^{1/4}$, expected from asymptotic solutions of Comisso et al. having a
  decreasing logarithmic dependence, and deduced from theoretical model of the ideal tearing mode proposed by
  Pucci \& Velli. Note that, only the regime where plasmoids can form is considered, as $S  \simgt S_c$ (with $S_c \simeq 5  \times 10^3$).
  Growth rate values using $\tau_A$ for normalization can be deduced as  $  \gamma_p  \tau_A  \simeq \gamma_p  t_A/2$.
  } 
\label{fig1}
\end{figure}
This initial setup is similar to the one used in the previously cited references \citep{ri90},
and rotated with an angle of $\pi/2$ compared to the
equilibrium chosen in the other studies \citep{kep14}.
Note that, the asymptotic (at large $r$) magnetic field strength 
is unity, and thus defines our normalisation. Consequently, our unit time in the following paper, will be defined as the Alfv\'en transit
time across the unit distance (i.e. the initial characteristic length scale of the dipole structure) as $t_A = 1$. The latter time
is slightly different from $\tau_A$ that is based on the half-length of the current sheet and on the upstream magnetic
field magnitude. However, in our simulations we can deduce that $\tau_A \simeq t_A/2$ (see below and Paper I).
In usual MHD framework using the flow velocity and magnetic variables, force-free equilibria using an additional
vertical (perpendicular to the $x-y$ plane) can be considered \citep{ri90}, or non force-free equilibria
can be also ensured trough a a thermal pressure gradient balancing the Lorentz force \citep{kep14}.
In our incompressible reduced MHD model, as thermal pressure is naturally absent, we are not
concerned by such choice. 

In previous studies using a similar physical setup, a square domain $[-R: R]^2$ was taken with $R$ large enough in order
to have a weak effect on the central dynamics. For example a standard value of $R = 3$ is taken in \citet{bat19}.
In the present work and in Paper I \citep{bat20}, a choice of using a circular domain with a radius $R = 3$ is done. We have checked that
it does not influence the results compared to the square domain setup. However, this allows the use of a lower number of
finite-element triangles (as the circle area is evidently smaller than the square for the same radius value $R$), and this
also simplifies the numerical boundary treatment as only one boundary instead of four in our finite-element
discretization are needed.

A stability analysis in the reduced MHD approximation using the energy principle has given that the linear
eigenfunction of the tilt mode is a combination of rotation and outward displacement \citep{ri90}. Instead of
imposing such function in order to perturb the initial setup, we have chosen to let the instability develops from
the initial numerical noise. Consequently, an initial zero stream function is assumed $ \phi_e (x, y) = 0$,
with zero initial vorticity $ \omega_e (x, y) = 0$.  The values of our four different variables are also
imposed to be constant in time and equal to their initial values at the boundary $r = R$.

\section{Results}
\begin{figure}
\centering
 \includegraphics[scale=0.8]{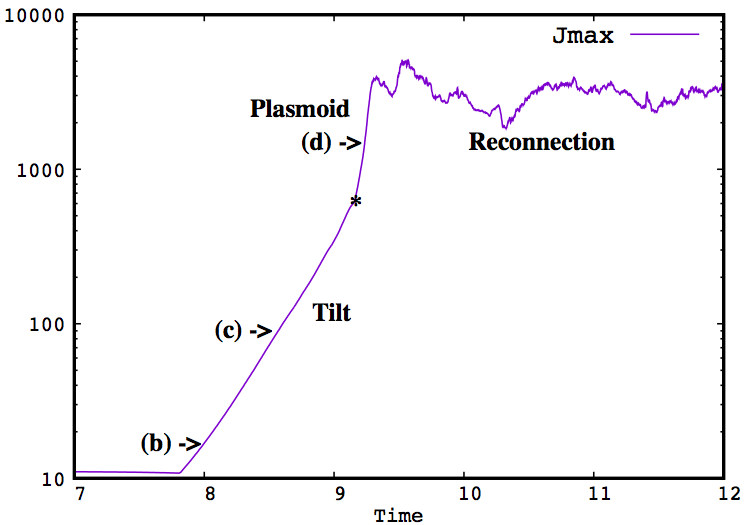}
  \caption {Time history of the maximum current density obtained for a run using $P_m = 1$ and $S^* = 1/\eta = 5 \times 10^4$
  (the corresponding Lundquist number is $S \simeq 1 \times 10^5$).
  The three different phases, namely the tilt development, the plasmoid chains formation, and the stochastic reconnection regime
  are indicated. The time is normalized using $t_A$ (see text).
  } 
\label{fig2}
\end{figure}

\begin{figure}
\centering
 \includegraphics[scale=0.2]{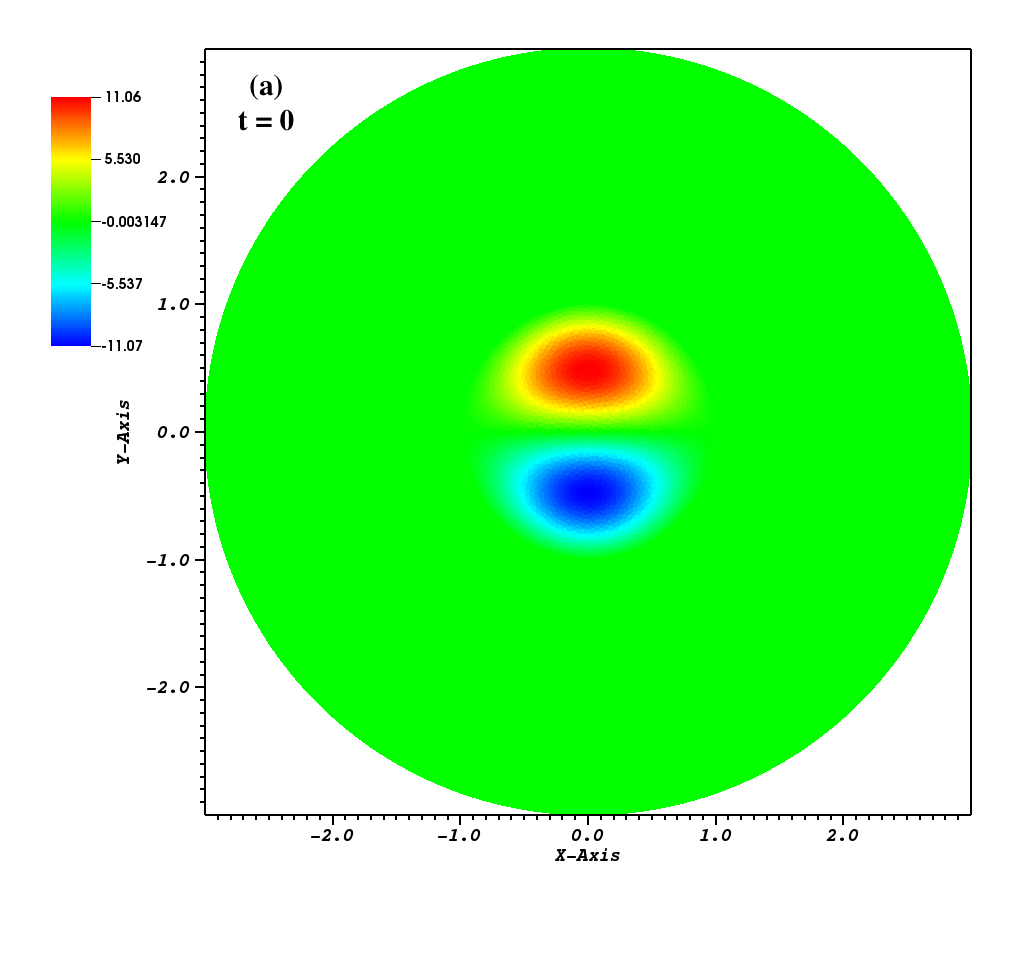}
 \includegraphics[scale=0.2]{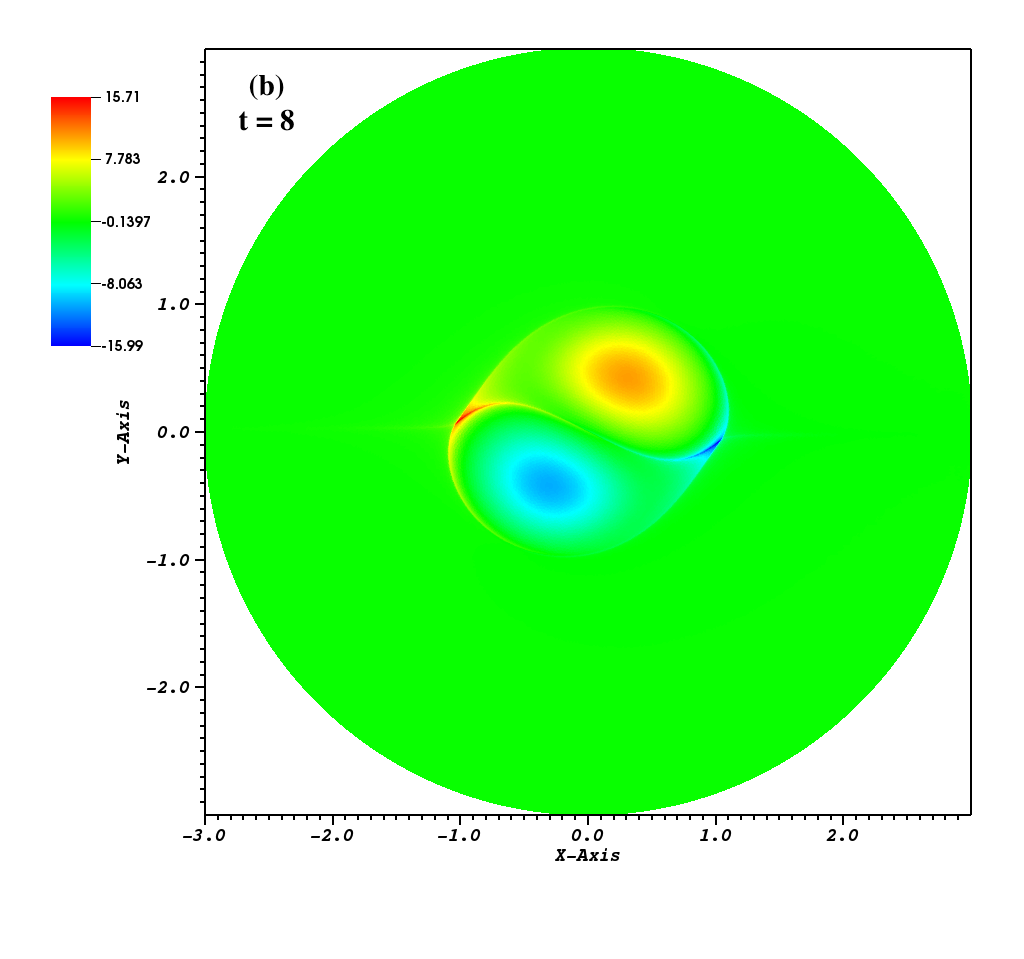}
  \includegraphics[scale=0.207]{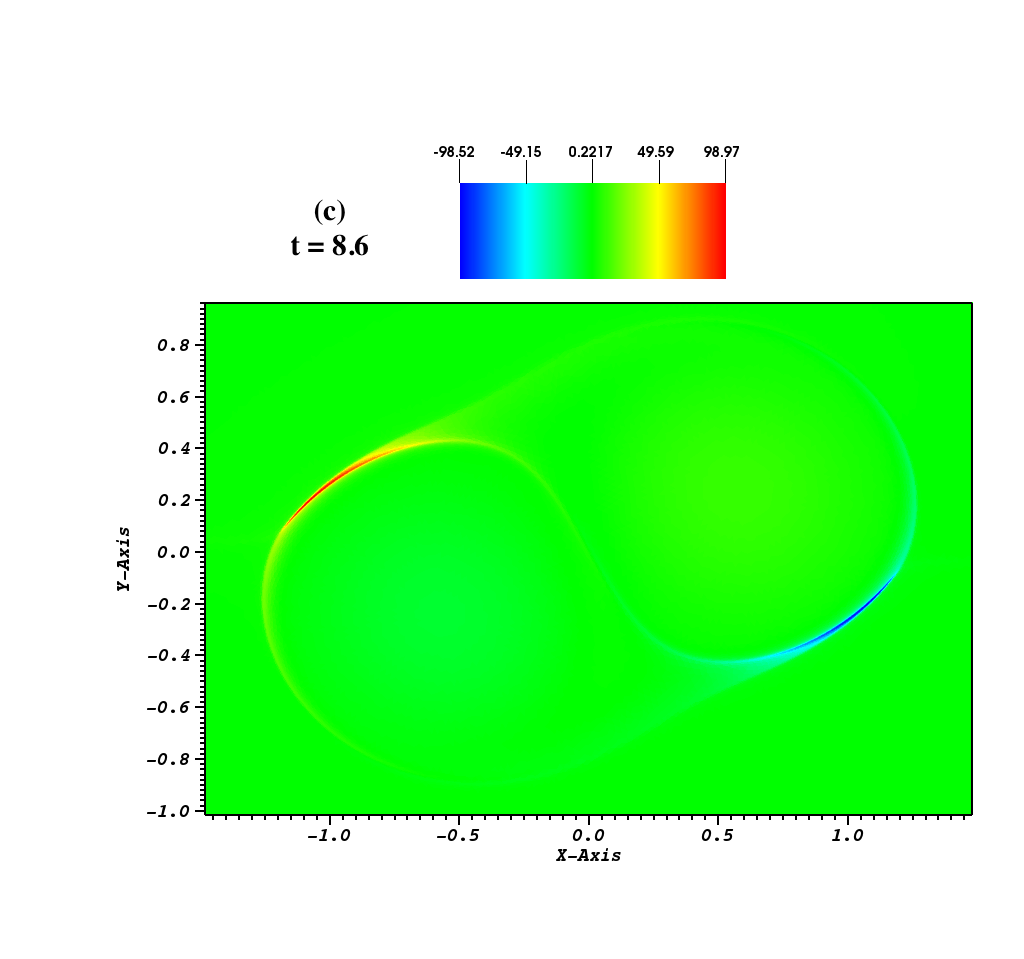}
  \includegraphics[scale=0.2]{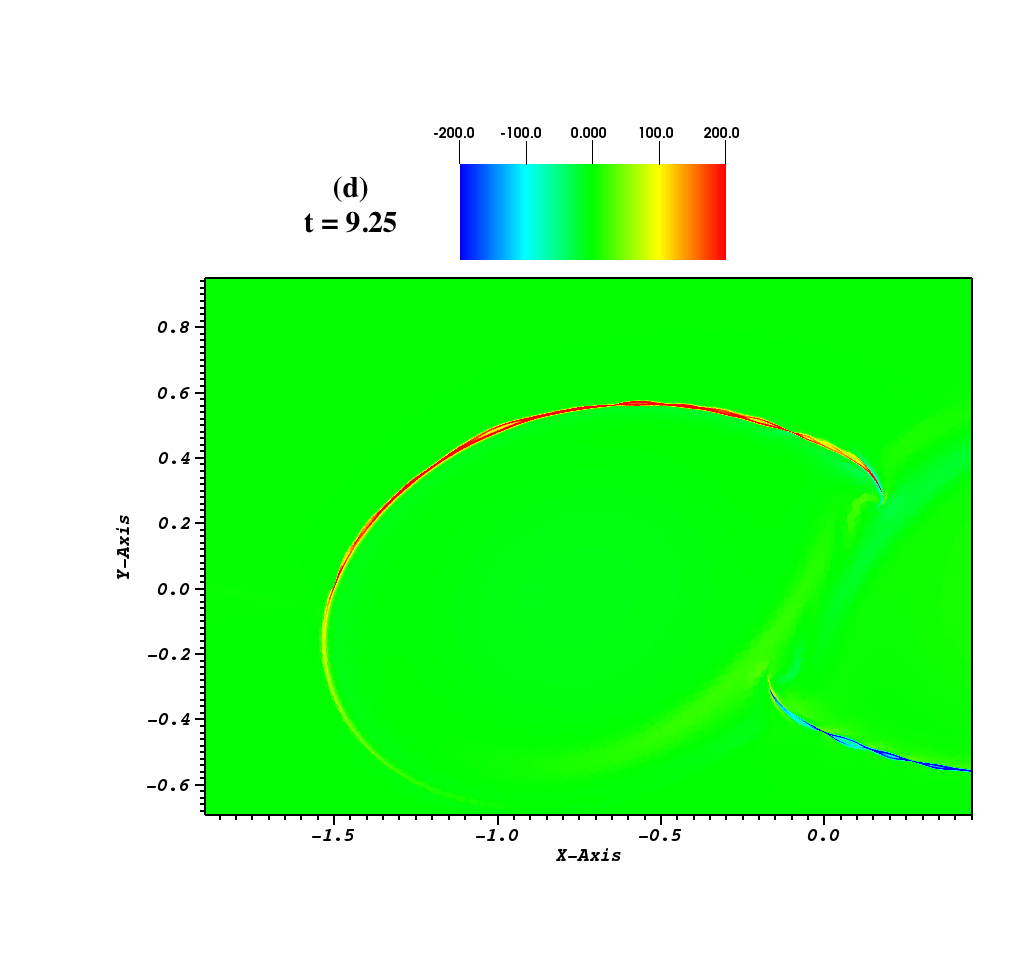}
  \caption {Snapshots of the colored contour map of the current density, corresponding to
  different times of the previous figure. Note that a zoom-in of the central region is used for (c)-(d), with
  additionally saturated values in the range $[-200:200]$ for (d) case.
  } 
\label{fig3}
\end{figure}

\subsection{Summary of the previous study at $P_m = 1$ (Paper I)}

First, from our knowledge, this study was the first one to address in detail the reconnection process associated with
the nonlinear evolution of the tilt instability. Interesting results, even in the SP (Sweet-Parker) regime were obtained. Indeed, two forming twin
current sheets (with current density of opposite sign) drive a steady-state reconnection in agreement with classical scaling laws
given by the famous Sweet-Parker model.  A slight amendment by a factor of two for the vorticity of the outflow is however required
due to the particular asymmetry associated with the curved geometry of the current layers (see Figure 8 in Paper I).

In this latter study using the tilt instability as a triggering mechanism to form the current sheets at magnetic Prandtl number $P_m = 1$,
the transition from a SP reconnection process to a plasmoid-dominated regime occurs for a critical Lundquist number
$S_c \simeq 5 \times 10^3$. This is a factor of two lower than the often-quoted $S_c \sim 10^4$ value in the literature. However,
there is no precise universal value, as it depends on different parameters like, the current sheet geometry (via the choice for the initial setup),
the magnetic Prandtl number, and also the noise amplitude (via the numerical scheme).
The exact definition of the Lundquist number can also differ slightly from one study to another. 
For coalescence instability, a value of $S_c \sim 3  \times 10^4$ has
been reported in simulations assuming zero explicit viscosity.
Even for Lundquist number very slightly lower than $S_c$, the formation of a transient single plasmoid
was observed to occur at a relatively late time, with no real impact on the SP reconnection rate (see Figures 9-10 in Paper I).

In the plasmoid-dominated regime, the growth of the plasmoids has been examinated, from their
birth to their ensuing disrupting effect on the current sheets. An important reference time scale for comparing the
latter growth is the time scale for forming the current layers, that is given by $\tau = 0.38$ $t_A \sim 1$ $\tau_A$, as the tilt
mode is an ideal MHD instability leading to a current density increasing exponentially as $e^{2.6 t}$ ($t$ being expressed in units of $t_A$).
This triggering phase in the simulations has been carefully checked to agree with stability theory
\citep{ri90}. As seen in Figure 6 of Paper I, the formation of the current sheets proceeds trough a combination
of thinning (as $a$ is observed to decrease in time), stretching (as $L$ is increasing), and a weak magnetic field 
strengthening, in agreement with the scenario suggested by \citet{tol18}.

In paper I, we have defined two simple parameters characterizing the growth of the forming plasmoids.
The first one, $t_p$, is the delay time between the birth of the first plasmoids (time at which they become
to be barely visible in the current density structure) and the start of the formation of the currents sheets (taken as the time at which the
corresponding current density exceeds the equilibrium setup value).
A rapidly converged constant value (with $S$) of $t_p = 2.4$ $\tau_A$ is obtained (see Figure 14 in Paper I).
This delay time has been identified to correspond to the quiescent phase proposed in Comisso et al.'s scenario, during which many modes become 
progressively unstable and compete with each other (see Figures 3-4 in \citet{com17}).
Indeed, the duration of this phase is predicted to be approximatively given
by the time scale of the current sheet formation. This is also in agreement with a conclusion drawn in \citet{uzd16}.
A similar result has been obtained for the coalescence setup, with a slight difference for the highest $S$ values
where their time delay is non-monotonic and increases weakly again \citep{hua17}. 
The second parameter is a growth rate, that is estimated by taking the second slope observed during the
increase of the maximum current density (see Figure 11 in Paper I), and thus characterizes an abrupt growth phase 
following the slower previous quiescent phase.
The latter growth rate was identified as $\gamma_p$, the growth rate of the dominant 
mode that emerges "first" at the end of the linear phase in the theory of \citet{com17}.

As one can see in Figure 1,  the results obtained for $\gamma_p$ in Paper I (using 
$P_m = 1$) qualitatively agree with a non-monotonic dependence with $S$,
as a consequence of the non-power law dependence with $S$ predicted by theory \citep{com17}.
Moreover, values $\gamma_p  \tau_A \simeq 10$ (as  $\gamma_p  t_A \simeq 20$) are also obtained 
for the highest $S$ values, thus confirming that $\gamma_p  \tau_A >> 1$ at the end of the linear phase.
For $S  \simgt S_c$, the scaling law given by SP stability theory with $\gamma_p  \propto S^{1/4}$ has been only marginally recovered.
A very similar result has been obtained for the coalescence setup \citep{hua17}. 
The difference at relatively low $S$ can be largely attributed to the reconnection outflow (neglected in theoretical models) that can affect the growth of the plasmoids
and thus the scaling relations \citep{hua19} As shown in previous studies, the noise induced by the numerical simulations also influences the results, and thus
is an important parameter that needs to be investigated in future studies.
Conversely, our results seem to contradict values deduced from the ideal tearing model of \citet{puc14}, where the linear growth of
plasmoids is predicted to be constant and at most Alfv\'enic (i.e. $\gamma_p  t_A \simeq 1$). 
Indeed, values of $\gamma_p  \tau_A \simeq 0.6$ and $\gamma_p  \tau_A \simeq 0.4$ are expected at
$P_m =0 $ and $P_m = 1$ respectively, assuming an Harris-type profile for the current layer. One must also note that
(as explained in introduction), this latter value is obtained by considering the aspect ratio $L/a$ to be equal to the critical value
$S^{1/3}$, that is in fact higher during our fast shrinking process of the current sheet (see Figure 16 in \citet{bat20}), consequently making possible a higher value.

The use of the time history of the maximum current density in order to estimate the growth rate of the plasmoids
in the simulations (as done in our previous study in Paper I) can be criticized. Indeed, we cannot rigorously prove that the relevant
phase (called plasmoid phase as one can see below) corresponds to an equivalent phase of linear development of tearing-type
instabilities taken from a theoretical stability study of a resistively unstable current layer. In other words, the comparison of
our estimated instantaneous growth rate (deduced from the current density time evolution) with the theoretical linear growth rate is not trivial.
In the context of the coalescence instability, \citet{hua17} separate the fluctuation magnetic field perturbation due
to the plasmoid instability from the background field contribution. And, the instantaneous growth rate was consequently derived from
the value of the perturbation as a function of time. The use of this technique (via a superposition of Chebyshev polynomials) is possible when the current layer
is straight. This is not the case in our study due to the curved nature of the current sheets (see Paper I and figures below in the present paper).
Nevertheless, we are able to give two strong arguments reinforcing the use of the maximum current density to estimate $\gamma_p$.
The first one consists of a close inspection of the current density structure during the plasmoid
phase, in order to check when non linear effects associated to the plasmoids growth (like coalescence for example) come into play.
The second one consists in doing additional runs at different magnetic Prandtl values, in order to examine the dependence of our
estimated $\gamma_p$ dependence with $P_m$ and compare with dependence predicted form visco-resistive linear theory. 

\begin{figure}
\centering
 \includegraphics[scale=0.4]{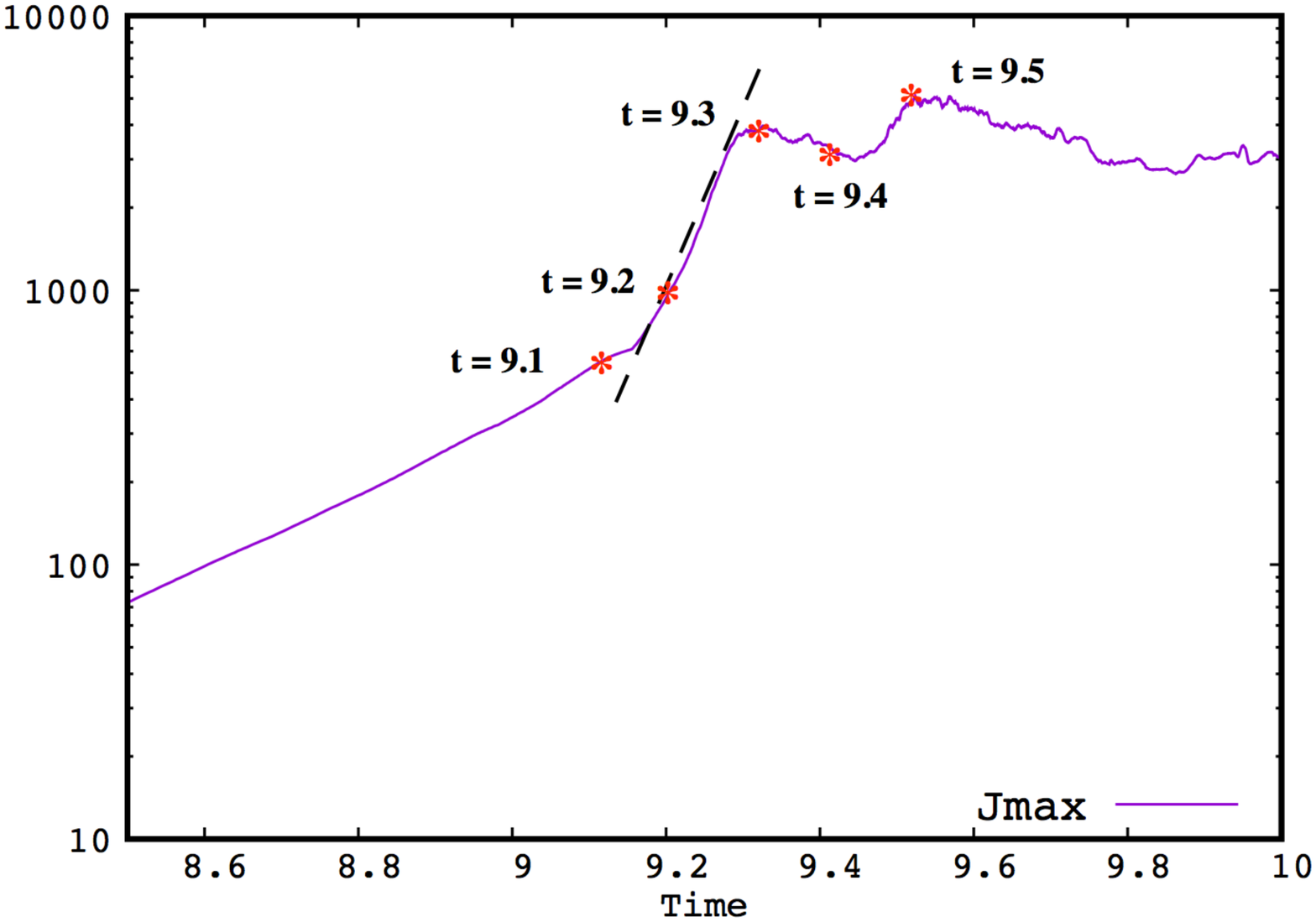}
  \caption {Zoom-in of the time history of the maximum current density obtained fo a run using $P_m = 1$ and $S^* = 1/\eta = 5 \times 10^4$,
  showing the transition between tilt, plasmoid, and reconnection phases.
   } 
\label{fig4}
\end{figure}

\begin{figure}
\centering
 \includegraphics[scale=0.22]{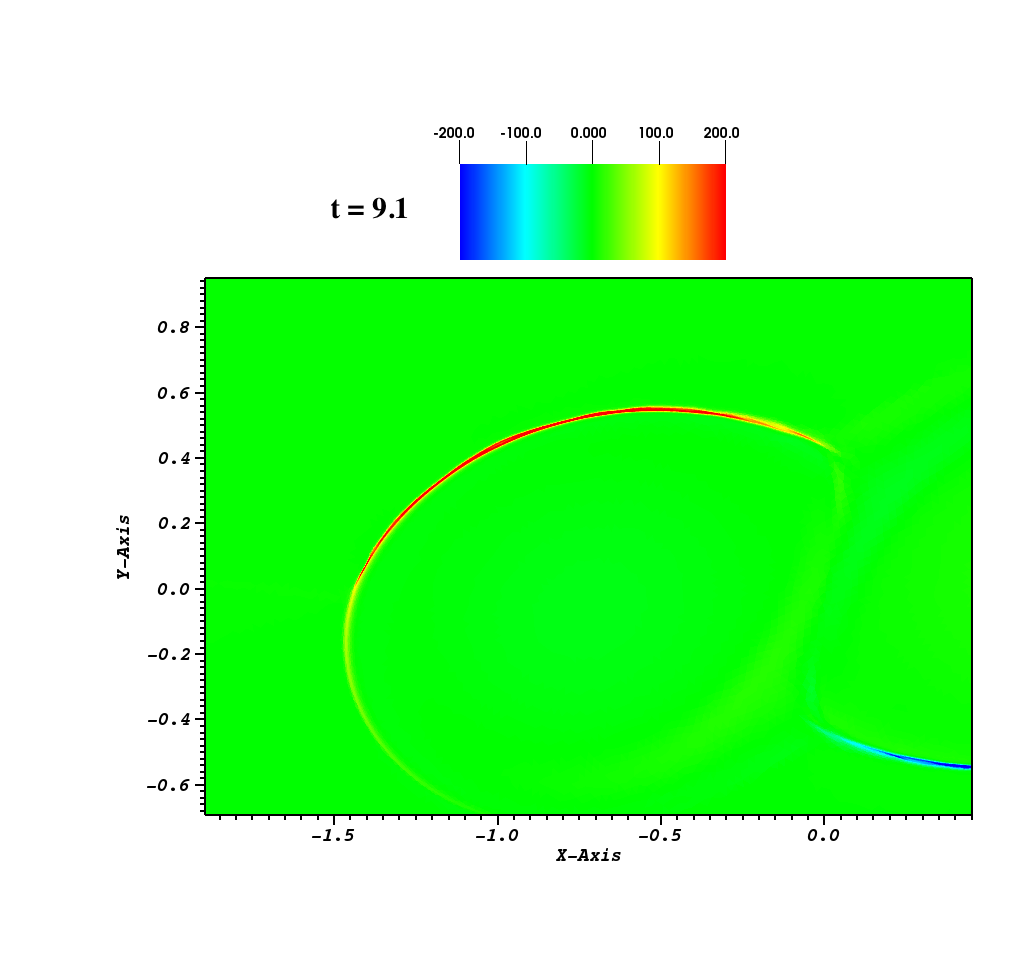}
 \includegraphics[scale=0.22]{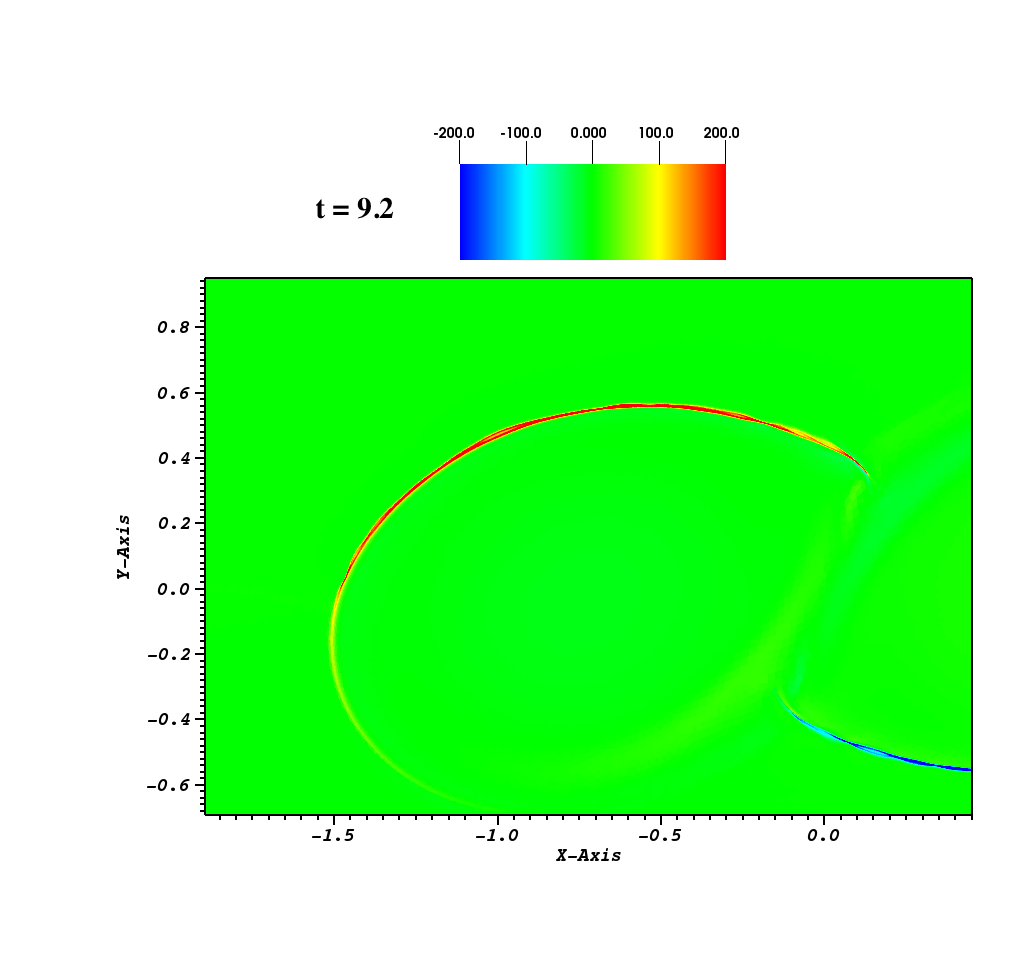}
 \includegraphics[scale=0.22]{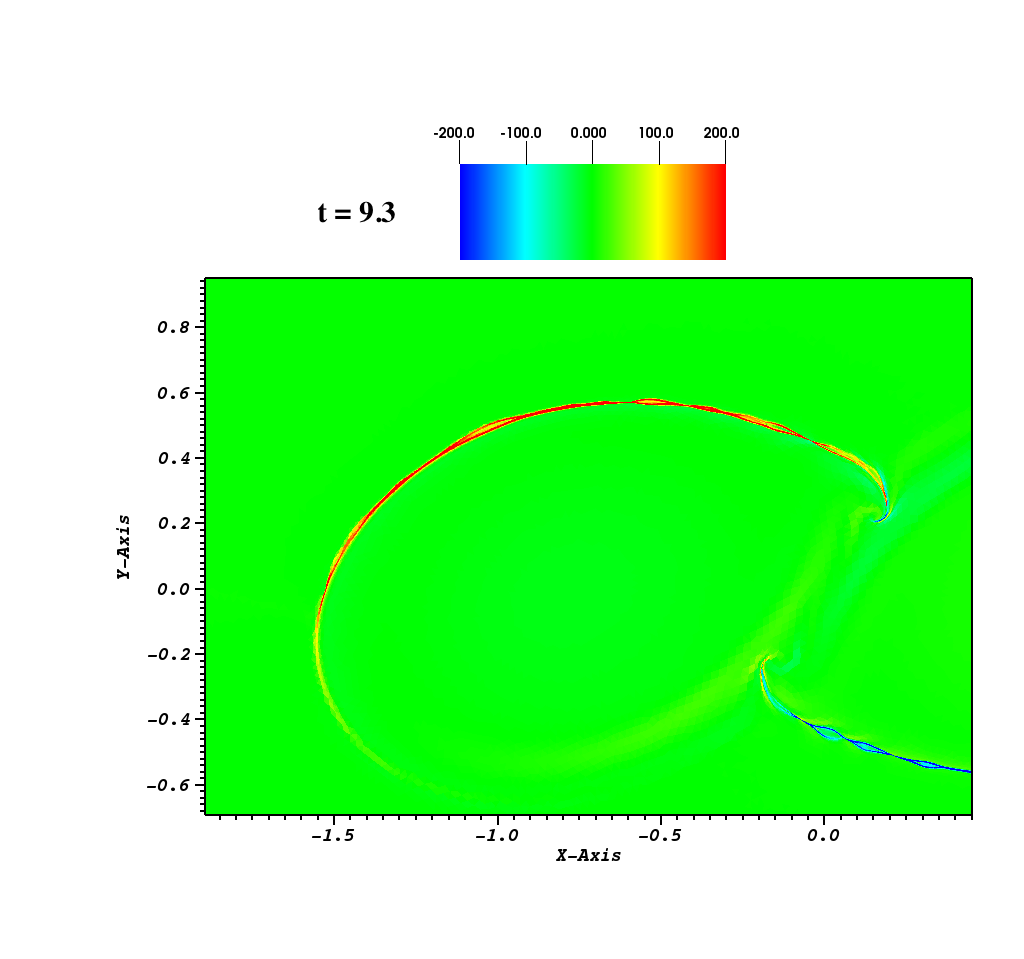}
 \includegraphics[scale=0.22]{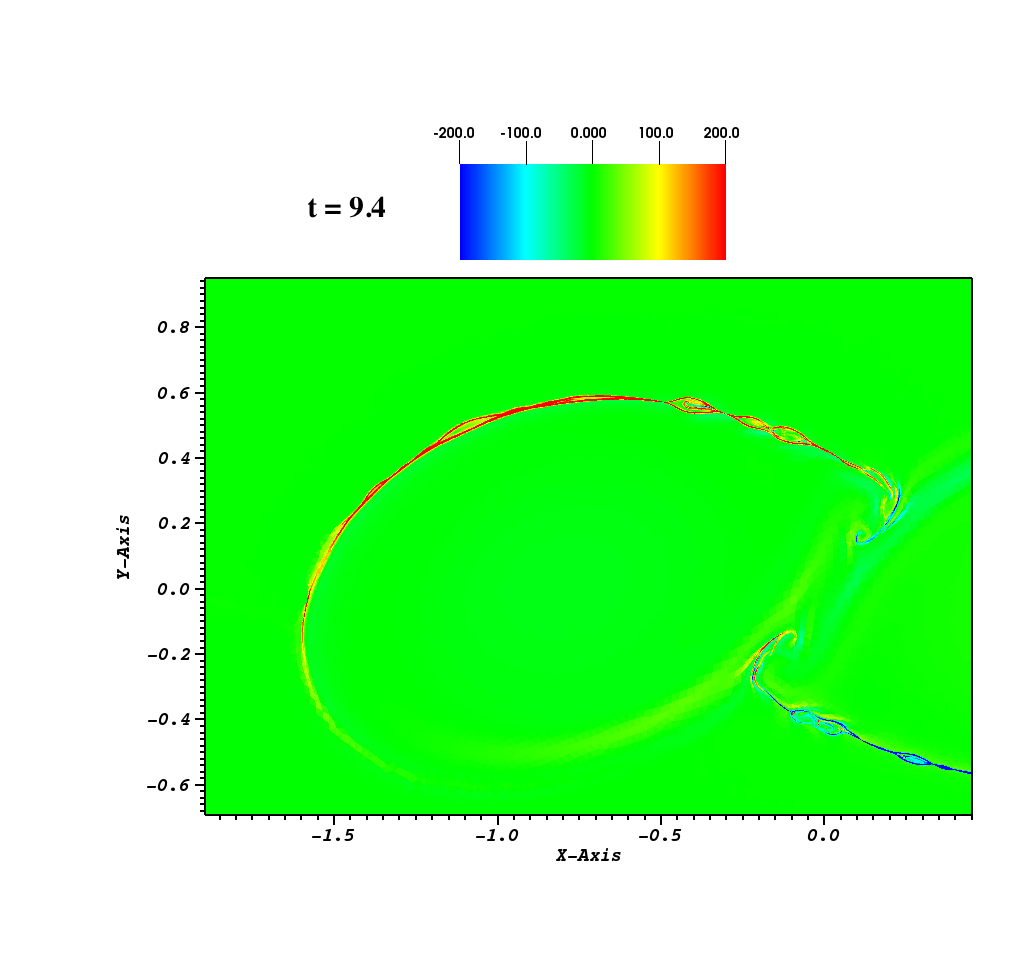}
  \caption {Snapshots of the colored contour map of the current density, corresponding to
  four different times of the plasmoid growth phase (see previous figure). Note that a zoom-in of the central region (centered on the left current layer) is used with
  additionally saturated values in the range $[-200:200]$.
   } 
\label{fig5}
\end{figure}

\begin{figure}
\centering
 \includegraphics[scale=0.22]{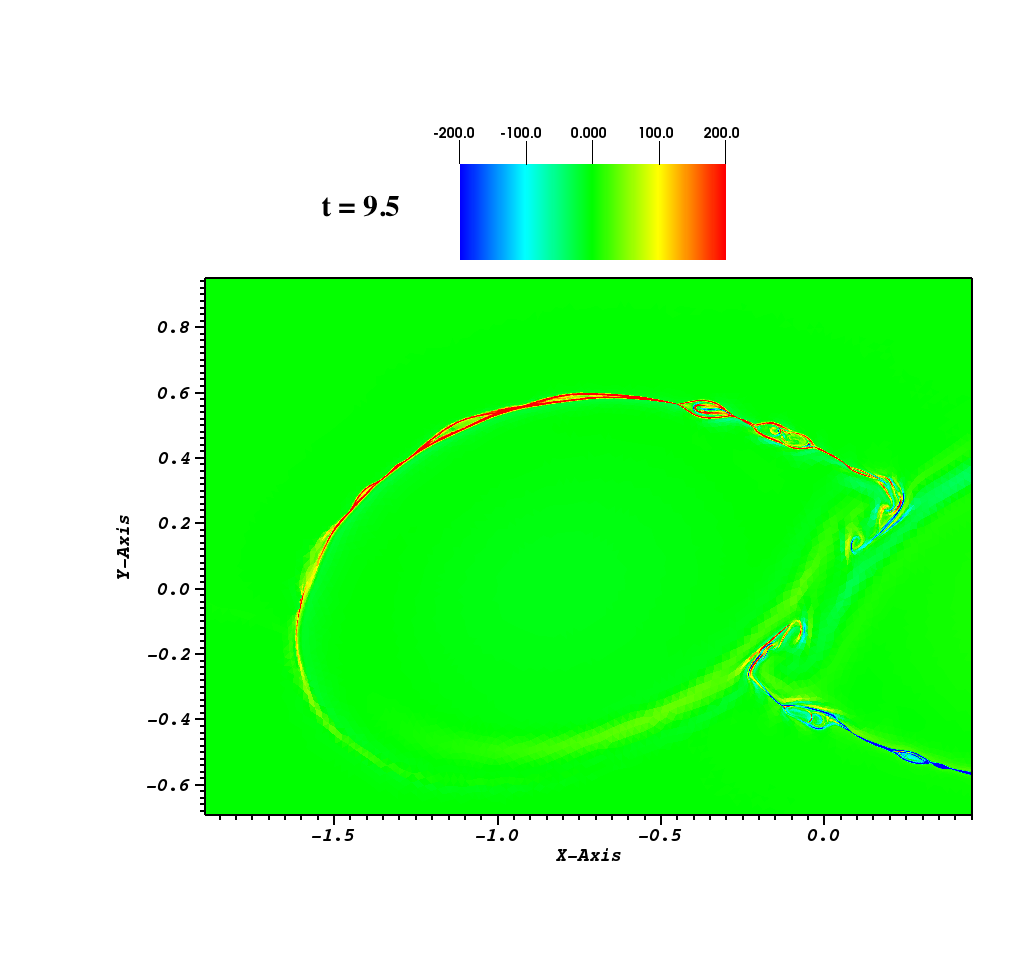}
 \includegraphics[scale=0.22]{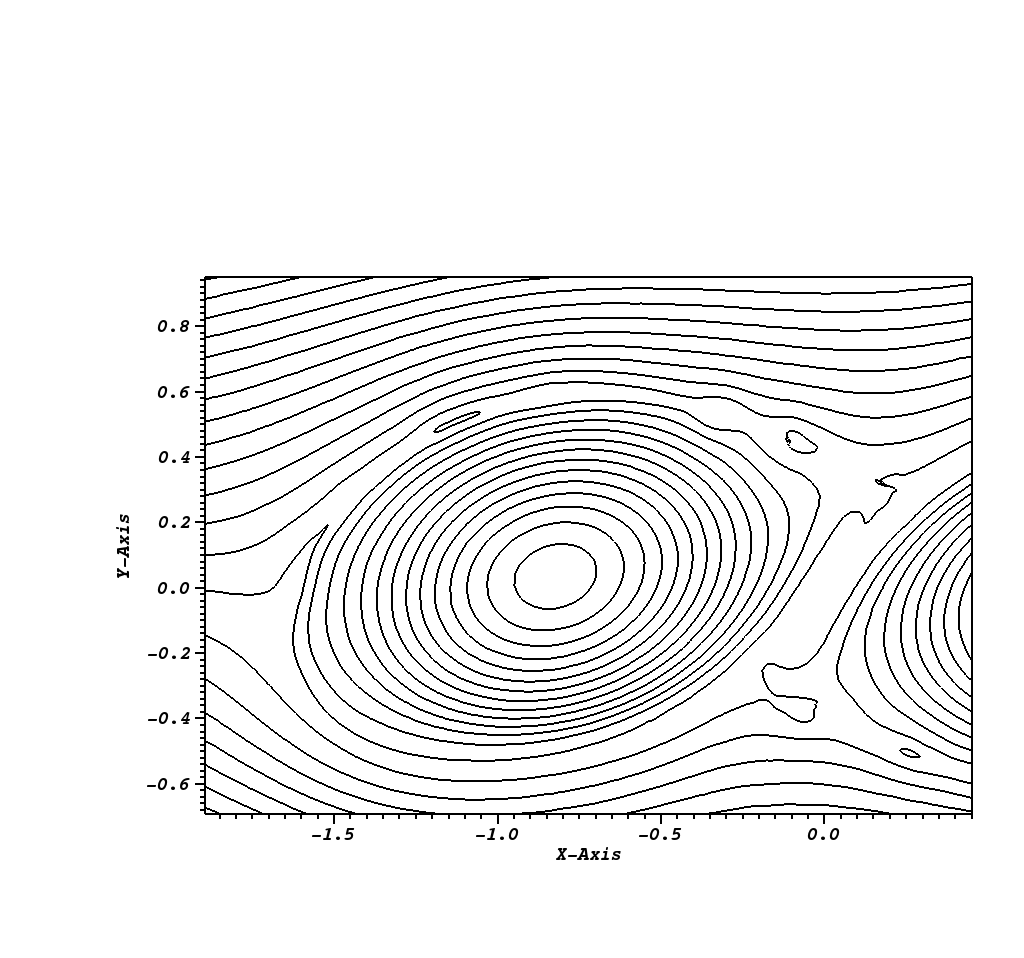}
   \caption { (Left panel) Snapshot of the colored contour map of the current density (centered on the left current layer), corresponding
    to an early time of the reconnection phase. (Right panel) Corresponding magnetic field lines obtained at the same time.
   } 
\label{fig6}
\end{figure}

\subsection{Detailed time history of the maximum current density obtained at $P_m = 1$}

First, in order to have an overview of the time evolution of the system, we have simulated a case using
$P_m = 1$ and $S^* = 1/\eta = 5 \times 10^{4}$. The corresponding Lundquist number $S = L V_A / \eta$ can be deduced by estimating
the half length of the current layer $L$ and the magnetic field $B_u$ (as $V_A$ is the Alfv\'en speed based
on the magnetic field amplitude in the upstream current layer $B_u$), leading to $S \simeq 10^{5}$.

The results obtained for the measured maximum current density (taken over the whole domain)
is plotted in Figure 2 as a function of time. One can see that, at $ t \simeq 7.8$ $t_A$, the tilt instability sets in as two twin current layers
are forming (see also Figure 3b). During the tilt phase, the intensity of the density current is increasing in time as $e^{2.6t}$,
as described in Paper I (see also Figure 3c for $t = 8.6$ $t_A$). As the tilt mode is an ideal MHD instability, this phase is not dependent of the
resistivity nor of the viscosity  \citep{ri90}. Later in the time evolution, an abrupt change of slope is clearly visible in Figure 2 
at $t \simeq 9.15$ $t_A$ (at time spotted by the first asterisk). During this slope increase, a chain of plasmoids progressively invades
each current layer (see Figure 3d at $t = 9.25$ $t_A$). The plasmoid phase typically ends when an oscillating quasi stationnary phase
is obtained with magnetic reconnection taking place (see later).

In paper I, the latter measured current density slope observed during the plasmoid phase was assumed to be a good estimate for the instantaneous growth rate of
the plasmoids. In order to check the validity of this assumption, a detailed time history of current layer structure is investigated
during the transition between these three phases. The results are plotted in Figures 4-6, for only one of the two current layers
for clarity. Indeed, at a time close to the transition between
the tilt and plasmoid phases, the plasmoids are barely visible. For example, at $t = 9.1$ $t_A$, a single plasmoid begins to appear at the
right corner of the current layer (see Figure 5a). Then, at  $t = 9.2$ $t_A$ (Figure 5b), other plasmoids begin to appear all along the layer.
At $t = 9.3$ $t_A$, the same plasmoids previously described have grown. Finally at $t = 9.4$ $t_A$, the plasmoids begin to coalesce (see right corner
in Figure 5d) indicating a non linear interaction for plasmoid dynamics. The latter coalescence is visible just after (see Figure 6a), and
magnetic reconnection is also at work at this time (Figure 6b).

We can conclude that during the plasmoid phase (i.e. between $t = 9.1$ $t_A$ and $t = 9.3$ $t_A$) used to determine the growth rate $\gamma_p$,
the structure of the current layers does not show any non linear behavior. Non linear effects (coalescence between primary islands) begin
to be visible only at $t \simeq 9.4$ $t_A$, thus validating our procedure.

\subsection{Simulations at different $P_m$ values}

The dependence of the plasmoid growth rate with the Prandtl number $P_m$ is investigated for two inverse resistivity values, taking
$S^* = 10^5$ and $S^* = 2 \times 10^4$ in the simulations. Note that, these two values of $S^*$ can be translated into the two corresponding Lundquist
number values, $S  \simeq 2 \times 10^5$ and $S  \simeq 4 \times 10^4$ respectively.
The results that are reported in Figure 7, clearly follow a fitted scaling law $\gamma_p  \propto (1 + P_m)^{-5/8}$. This is in agreement with 
predictions from linear theory, as for example $\gamma_p  \tau_A  \simeq 0.62$  $(1 + P_m)^{-5/8} S^{1/4}$ derived
in \citet{com16}.
Consequently, it is very unlikely that the second slope increase of the maximum current density is a non linear effect.
Otherwise, it would give another $P_m$ dependence with a probably less sensitivity to viscosity.

\begin{figure}
\centering
 \includegraphics[scale=0.82]{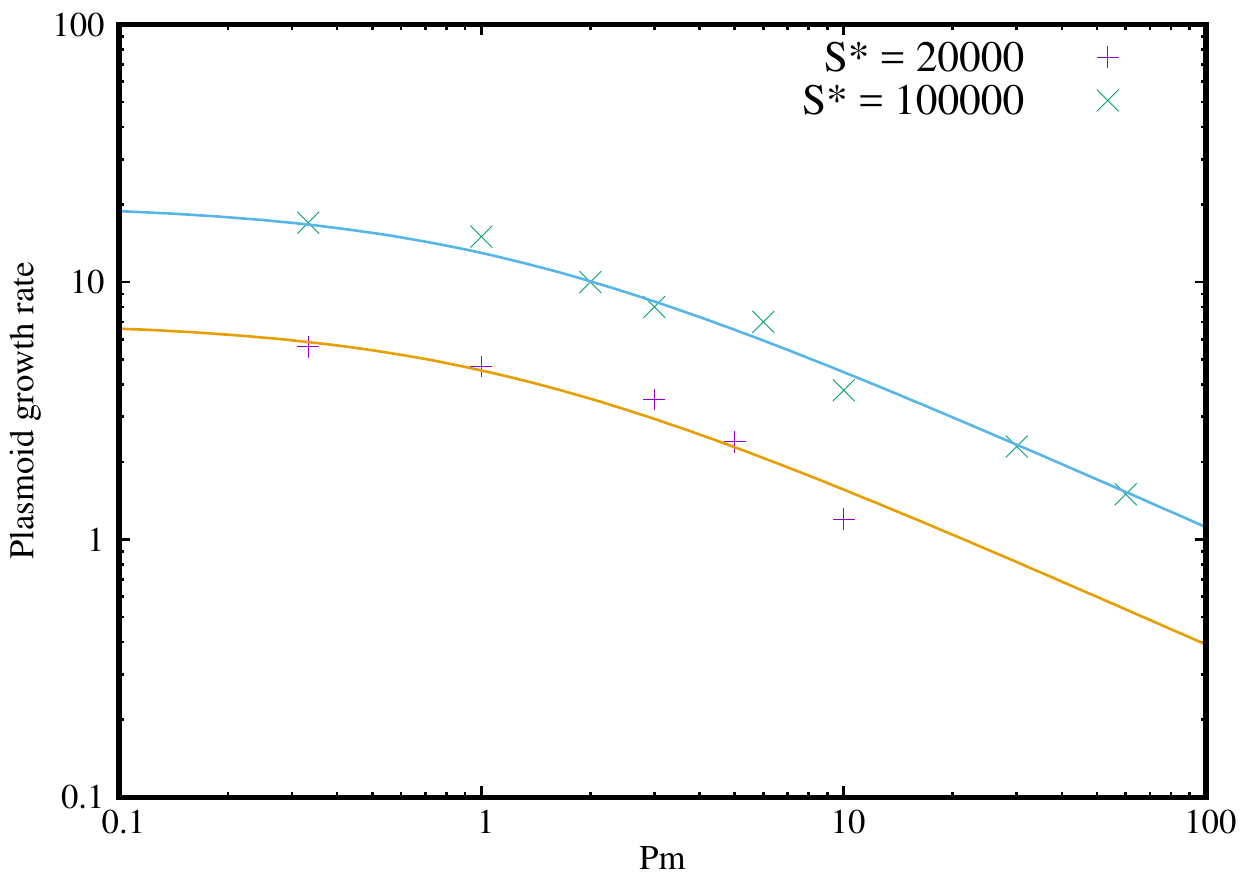}
   \caption {Plasmoid growth rate normalized to $t_A^{-1}$, i.e. $\gamma_p t_A$, obtained in different runs as a function
   of the magnetic Prandtl number $P_m$, for two inverse resistivity values ($S^*$ = $2 \times 10^4$ and $1 \times 10^5$). The theoretical dependences (see text) scaling as
   $(1 + P_m)^{-5/8} $ are plotted for comparison.
   } 
\label{fig7}
\end{figure}

\
\section{Conclusion}

In this work, we have extended a previous study (see Paper I) devoted to the formation of chains of plasmoid during magnetic
reconnection in the 2D MHD framework. More precisely, the focus was on addressing the onset phase in relation with
the linear stability theory in forming quasi-singular current layers, in the plasmoid dominated regime for which the Lundquist
number is higher than the critical value $S \simgt S_c$. Numerical simulations with FINMHD code are carried out at different
magnetic Prandtl values in this high Lundquist number limit. The tilt mode is used as the initial setup to form the current layers
on a fast ideal MHD time scale. 

Our results confirm that, an onset phase characterized by a sudden super Alfv\'enic growth of plasmoids is obtained
(with $\gamma_p \tau_A \simeq 10$ for our runs), as predicted by the stability theory proposed Comisso and collaborators \citep{com16, com17}.
The simple diagnostic using the time evolution of maximum current density is checked to be valid for this aim.
During this phase, the plasmoids remain in a linear growth regime, and the transition to a non linear regime occurs when
the statistical steady-state with oscillating current density is reached. This latter phase is characterized by a time-averaged
reconnection rate nearly independent of the Lundquist number (see Paper I).
Our results being very similar to the those obained from the coalescence setup, suggest that Comisso et al.'s model
is able to correctly predict the explosive growth of plasmoids leading to disruption of the reconnection current sheets when the
initial configuration is ideally unstable. On the other hand, the other model developped by \citet{puc14}, where the plasmoid linear
growth is at most Alfv\'enic, could apply when the
initial configuration is ideally stable (and thus resistively unstable), as it has been validated using Harris-type current layer.
This could explain the fastest time scale involved in the first category of setup (ideally unstable) compared to the second one (ideally
stable).

The time-averaged normalized reconnection rate reported in Paper I is $0.014$, that is two times higher than the value deduced
from the coalescence setup. It is nevertheless in good agreement with values obtained in the literature of $\sim 0.01$ much higher
than the Sweet-Parker rates, which could be sufficient to explain many disruptive events if the collisionnal regime apply.
A fractal model (with hierarchical structure of the plasmoid chains that are effectively observed in simulations) based on heuristic arguments
has been proposed to explain this fast rate independent of the Lundquist number \citep{hua10}. Indeed, to this end, a number of plasmoids
(called non linear number) is required to scale linearly with $S$ \citep{hua10}. Investigating this point is a complicated task
requiring longer time simulations, and it will be the subject of a future study using tilt setup.

The Lundquist number reached in this study is high enough in order match the relevant values for
tokamaks. Indeed, the relevant $S$ value for the internal disruption associated with the internal kink mode is
$S  \simeq 10^5$, as $S = 0.004 S^*$ ($S^* = 2.5$ $ 10^7$ being a standard Lundquist number value defined
in terms of the toroidal magnetic field)  \citep{gun15}. The corresponding width of the Sweet-Parker current layer is thus estimated to be $a \simeq 1$ cm,
and the smallest length scale associated to the plasmoid structure is probably of order $1$ mm or even smaller, reaching
thus a scale close to the the kinetic ones. Kinetic effects could be incorporated to our model in order to address this
point. For example, the plasmoid instability has been shown to facilitate the transition to a Hall reconnection 
in Hall magnetohydrodynamical framework with an even faster reconnection rate of $\sim 0.1$ \citep{hua11}.
The smallest length scale associated to the plasmoid structure for $S = 10^6$ remains larger than the kinetic scale 
that is of order $10$ m, when considering a solar loop structure and taking a length $L = 10^7$ m. However, as very high
Lundquist number (at least $10^{10}$) is required for the solar corona, kinetic effects could also play a role if the kinetic scale
is reached via the plasmoid cascade at such huge Lundquist number.

% susie put cite commands here, don't bother with citet etc just yet.

\bibliographystyle{jpp-tilt}
% Note the spaces between the initials

\end{document}